\documentclass[12pt]{aastex}
\renewcommand\th{\thinspace}
\newcommand\kms{\ifmmode{\rm km\th s^{-1}}\else km\th s$^{-1}$\fi}
\newcommand\cmss{\ifmmode{\rm cm\th s^{-1}}\else cm\th s$^{-2}$\fi}
\newcommand\msun{\ifmmode{M_{\odot}}\else $M_{\odot}$\fi}
\newcommand\rsun{\ifmmode{R_{\odot}}\else $R_{\odot}$\fi}
\begin{document}
\title{ Sub-Subgiants in the Old Open Cluster M67?}
\author{ Robert D. Mathieu}
\affil{Department of Astronomy, University of Wisconsin--Madison, 475 
  North Charter Street, Madison, WI 53706-1582}
\email{mathieu@astro.wisc.edu}
\author{Maureen van den Berg}
\affil{Osservatorio Astronomico di Brera, Via E. Bianchi 46, 23807 Merate (LC) Italy}
\email{vdberg@merate.mi.astro.it}
\author{Guillermo Torres and David Latham}
\affil{Harvard-Smithsonian Center for Astrophysics, 60 Garden Street, Cambridge, MA  
02138}
\email{gtorres@cfa.harvard.edu; dlatham@cfa.harvard.edu}
\author{Frank Verbunt}
\affil{ Astronomical Institute, Utrecht, The Netherlands}
\email{F.W.M.Verbunt@phys.uu.nl}
\author{Keivan Stassun}
\affil{Department of Astronomy, University of Wisconsin--Madison, 475 
  North Charter Street, Madison, WI 53706-1582}
\email{keivan@astro.wisc.edu}
\begin{abstract}
We report the discovery of two 
spectroscopic binaries in the field of the old open cluster M\,67 -- S\,1063 and S\,1113 -- whose
positions in the color-magnitude diagram place them $\approx$ 1 mag 
below the
subgiant branch. A ROSAT study of M\,67
independently discovered these stars to be X-ray sources. Both 
have proper-motion membership probabilities
greater than 97\%; precise center-of-mass velocities are consistent
with the cluster mean radial velocity. S\,1063 is also projected
within one core radius of the cluster center.   S\,1063 is a
single-lined binary with a period of 18.396 days and an
orbital eccentricity of 0.206. 
S\,1113 is a double-lined system with a circular
orbit having a period of 2.823094 days. 
The primary stars of both binaries are subgiants. The secondary
of S1113 is likely a 0.9  $M_{\odot}$ main-sequence star, which
implies a  1.3  $M_{\odot}$ primary star. We have been unable to
explain securely the low apparent luminosities of the primary stars.
The colors of S1063 suggest 0.15 mag higher reddening than found for either
M67 or through the entire Galaxy in the direction of M67. S1063 could be
explained as an extincted M67 subgiant, although the origin of such 
enhanced extinction is unknown.
The photometric properties of S1113 are well modeled by a cluster binary with a 0.9  $M_{\odot}$ main-sequence
secondary star. However, the low composite luminosity requires a small (2.0 $R_{\odot}$) primary star that
would be supersynchronously rotating, in contrast to the short synchronization timescales, the circular orbit, and the 
periodic photometric 
variability with the orbital period. Geometric arguments based on a tidally
relaxed system
suggest a larger (4.0 $R_{\odot}$) primary star in a background binary, but
such a large star violates the observed flux ratio.
Thus we have not been able to find a compelling solution for the
S1113 system.
We speculate that S1063 and S1113 may be the products of close stellar encounters involving binaries
in the cluster environment, and may define
alternative stellar evolutionary tracks associated with mass-transfer episodes, mergers, and/or 
dynamical stellar exchanges.
\end{abstract}
\section{Introduction}
The open cluster M\,67 is one of the most comprehensively studied of
all star clusters, and has long been the prototype for old (4 Gyr)
open clusters in the Galaxy. Indeed, one of the first photoelectric
color-magnitude diagrams was derived for M\,67 by Johnson \&\ Sandage
(1955).  Since that time the precision of stellar photometry has
steadily improved, and with it has the definition of the M\,67 giant
branch (Janes \&\ Smith 1984, Montgomery et al. 1993). Indeed the
remarkable narrowness of the M\,67 giant branch has served as a precise
touchstone against which innumerable single-star evolution models have
been tested (e.g., Dinescu et al. 1995).

However, even after application of strict proper-motion membership
criteria, the color-magnitude diagram of M\,67 remains littered with
stars that do not fall on a single-star isochrone (Fig.~\ref{s1063:cmd}). Some
of these seeming anomalies can be accounted for. For example, one
luminous star to the blue of the giant branch is a spectroscopic
binary whose composite light can be explained by a
giant--main-sequence pairing; another is a giant--white dwarf pair
with a complicated history (Mathieu et al.  1990, Verbunt \&\ Phinney
1995, Landsman et al. 1997). Many of the stars immediately above the
turnoff and subgiant branch are spectroscopic binaries and
consequently overluminous. And the parallel sequence of stars to the
red of the main sequence is assuredly comprised of 
binaries as well (Montgomery et
al. 1993).

Still, there remain stars that are not so easily explained. Most
famous are the blue stragglers, first noted in M\,67 by Johnson \&\
Sandage. The origin of blue stragglers in an open cluster environment
is still not securely understood today (Bailyn 1995, Hurley et al. 2001). 
Similarly, the
yellow giant S\,1072, lying 1.2 magnitudes in V above the main-sequence
turnoff, has defied explanation. A kinematic member in all three
dimensions and lying in projection in the cluster core, its explanation
as a non-member remains possible but not satisfying (Mathieu \&\
Latham 1986, Nissen et al. 1987). In this paper we consider two more
stars -- S\,1063 and S\,1113 -- which by every indication are cluster
members, yet whose location in the M\,67 color-magnitude diagram is
dramatically inconsistent with single-star evolutionary
theory. Specifically, as shown in Fig.~\ref{s1063:cmd}, these stars lie below
the subgiant branch.

Our attention has converged on these stars from two directions. We
have underway a multi-decade survey of the spectroscopic binary
population in M\,67 (Mathieu et al. 1990, Latham et al. 1997). 
Both S\,1063 and S\,1113 were found to be spectroscopic
binaries, with S\,1113 in particular being notable for the rapid
rotation of its primary star. Belloni et al. (1998) have also undertaken a
comprehensive study of stellar X-ray sources in M\,67 based on ROSAT
observations. S\,1063 and S\,1113 were
independently discovered to be X-ray sources, of which there are only
25 known among cluster members. Both stars are also photometric
variables (van den Berg et al. 2002), which are infrequent
in a cluster of this age (Stassun et al. 2002).  

In this paper we present a comprehensive discussion of the
orbital, spectroscopic, and photometric properties of S\,1063 and
S\,1113. Regrettably, like the blue stragglers and S\,1072, their
interpretation remains a puzzle.

\section{The stars}

The stars S\,1063 and S\,1113 had attracted some attention prior to
the radial-velocity and X-ray studies. Both stars are included in the
highly precise proper-motion study of Girard et al.  (1989), in which
S\,1063 is given a proper-motion  membership probability of 97\% and
S\,1113 of 98\%.  Being a bit less than one core radius from the
cluster center, S\,1063 has been included in most photometric studies
of M\,67; S\,1113 is located at three core radii, and thus only has
been observed in the wider-field photometric surveys. Racine (1971)
noted that S\,1063 was photometrically unusual in that it lies roughly
a magnitude in $V$ below the subgiant branch. S\,1113 has
photometric properties very similar to those of
S\,1063, but seems to have escaped notice
until mentioned by Kaluzny \&\ Radczynska (1991). In Fig.~\ref{s1063:cmd} we
show the color-magnitude diagram of M\,67 derived from the photometry
of Montgomery et al. (1993); S\,1063 and S\,1113 are indicated with
boxes.

Both stars have been found to be photometric variables. Racine (1971)
noted a large range (0.18 mag) in photometric observations of S1063,
and its variability was subsequently confirmed by other observers
(Rajamohan et al. 1988, Kaluzny \&\ Radczynska 1991). S\,1113 has a
variable star designation -- AG Cnc -- and Kaluzny \&\ Radczynska
(1991) found S\,1113 to be photometrically variable at the 0.05 mag
level. Based on the variability (but not light curves) of the stars 
and their
location in the color-magnitude diagram, Kaluzny \&\ Radczynska
(1991) suggested that both stars are highly evolved W UMa binaries
with extremely small mass ratios and consequently small photometric
amplitudes.

Both S\,1063 and S\,1113 were included in a large survey for
spectroscopic binaries among M\,67 proper-motion cluster members
begun in the mid-1980's (last summarized in Latham et al. 1997).
S\,1063 was found to be a single-lined spectroscopic binary,
with no evident distinction otherwise. S\,1113 was immediately
recognized as an unusual double-lined spectroscopic binary in that
the primary star was a rapid rotator.

Independently, ROSAT PSPC observations of M\,67 identified these two
stars as among 25 cluster members detected to have X-ray emission
(Belloni et al. 1993, 1998). As members of this select group, S\,1063
and S\,1113 received optical spectroscopic attention by Pasquini \&\
Belloni (1998) and van den Berg et al. (1999). Both stars
show strong emission cores in the Ca II H and K lines, indicative of
chromospherically active stars. Both stars also show H$\alpha$
emission. In S1063 the H$\alpha$ line is asymmetric, showing emission
which is blue-shifted with respect to an absorption feature. The position of the
absorption line agrees with the velocity of the primary, perhaps
suggesting that the absorption can be attributed to the primary star and that the
H$\alpha$ emission originates elsewhere in the system. Alternatively,
the H$\alpha$ line profile is very similar to that seen from HK\,Lac
by Catalano \&\ Frasca (1994), who attribute the emission to a large
flare lasting 6 days. In S\,1113 the H$\alpha$ emission is broad and
clearly centered on the velocity of the primary star, with an
equivalent width of 15 \AA.

\section{Observations and data analyses}

\subsection{Speedometry and orbital solutions}\label{s1063:speed}
Radial-velocity observations of S\,1063 were begun in
1987, and 28 observations were obtained through 1989. 18
radial-velocity observations were obtained for S\,1113 between 1989
and 1998. All observations were obtained with the Center for
Astrophysics (CfA) Digital Speedometers (Latham 1992). Two nearly
identical instruments were used on the Multiple Mirror Telescope 
\footnote {Some of the observations
reported here were obtained with the Multiple Mirror Telescope, a joint facility of the
Smithsonian Institution and the University of Arizona.}
and
the 1.5-m Tillinghast Reflector atop Mt. Hopkins, Arizona.  Echelle
spectrographs were used with intensified photon-counting Reticon
detectors to record about 45 \AA\ of spectrum in a single order near
5187 \AA\ with a resolution of 8.3 km s$^{-1}$ and signal-to-noise
ratios ranging from 8 to 15 per resolution element.

The 28 radial velocities for S\,1063 were measured using the
one-dimensional correlation package {\sc rvsao} (Kurtz \&\ Mink 1998)
running inside the IRAF \footnote{ IRAF is distributed by the National Optical 
Astronomy Observatories, which is operated by the Association of Universities for Research in Astronomy, Inc., 
under contract with the National Science Foundation.}
 environment. The template spectrum was drawn
from a new grid of synthetic spectra (Morse \&\ Kurucz, in
preparation) calculated using Kurucz's code ATLAS9. We correlated our
observed spectra against an array of solar-metallicity templates
spanning effective temperature, surface gravity, and projected
rotation velocity. The template giving the highest peak correlation
(averaged over all observations) had $\log g = 3.5$, $T_{\rm eff} =
5000$ K, and $v\sin i$ = 6 km s$^{-1}$. The radial velocities derived
with this template are presented in
Table~\ref{s1063:rv1063t}. The average internal error of these observations
is 0.86 km s$^{-1}$.

A single-lined orbital solution with period $P = 18.396$ days is
easily derived from these data; the elements are given in
Table~\ref{s1063:rv1063el} and the theoretical velocity 
curves are superimposed on
the data in Fig.~\ref{s1063:rv1063f}. The center-of-mass velocity of
$\gamma$ = 34.3 $\pm$ 0.2 km s$^{-1}$ is consistent with the cluster
mean radial velocity of 33.5 km s$^{-1}$ and observed velocity
dispersion of 0.5 km s$^{-1}$ (Mathieu 1983), so the star is a
kinematic cluster member in all three dimensions. Otherwise, the
orbital elements are unremarkable.

The analyses of the S\,1113 spectra were done using the 
two-dimensional correlation technique TODCOR
(Zucker \&\ Mazeh (1994), as implemented at CfA). A
two-dimensional array of templates in effective temperature and
projected rotation velocity, again derived from Kurucz spectra, were
correlated with all of the S\,1113 spectra. The effective temperatures
and rotation velocities of the templates were selected so as to 
maximize the correlation peak heights, averaged over all 18
spectra and weighted according to signal-to-noise ratio. 
We find projected rotational velocities $v\sin i$ of 53 km
s$^{-1}$ for the primary and 11 km s$^{-1}$ for the secondary,
respectively, with a 1-$\sigma$ uncertainty of 2 km s$^{-1}$ on
each. Van den Berg et al. (1999) previously measured
$v\sin i$ for both stars, finding 45$\pm$6 km s$^{-1}$ for the primary
and 12$\pm$1 km s$^{-1}$ for the secondary. Our derived effective
temperatures are 4800 K and 5500 K for the primary and secondary,
respectively, with a 1-$\sigma$ uncertainty of 150 K. Finally, we
derive a luminosity ratio at 5187\AA\ of 0.35$\pm$0.02
(secondary/primary). With color corrections we find a luminosity ratio at V
of 0.32 or 1.23 mag.

The primary and secondary radial velocities derived with these
template parameters are given in Table~\ref{s1063:rv1113t}. We obtained
a double-lined orbital solution using equal weighting for all velocities.
If the eccentricity is allowed to vary, the best-fit value is
$e = 0.022\pm0.010$. As this measure is not significantly different from
zero, we present elements for a circular (e=0) orbital solution
in Table~\ref{s1063:rv1113el} and show the orbit curves
superimposed on the phased data in Fig.~\ref{s1063:rv1113f}. (In these elements, $T_0$ represents a time of 
maximum radial velocity.) Again, the
center-of-mass velocity of $\gamma$ = 33.4$\pm$0.4 km s$^{-1}$ is consistent
with the cluster mean radial velocity. The mass ratio 
M$_2$/M$_1$ of the system is $q = 0.703 \pm 0.012$.

\subsection{Photometry and variability}

Van den Berg et al. (2002) have
photometrically monitored both stars, with 1-m telescopes at ESO (La
Silla), Kitt Peak and La Palma, in order to look for light variations
that may provide insight into the nature of the systems. We here
summarize their main results.

Van den Berg et al. (2002; Figure 5) also find S\,1063 to be variable, 
but as yet no period has been
identified in the photometric data.  The variability is clearly of a long-term
nature. If it is periodic, the 18-day-long ESO light curve establishes that
the period is longer than the pseudo-synchronous period of 14.64 days (Hut 1981, eq.42; see also Claret
\& Gim\'enez 1993). Furthermore, the ESO and Kitt Peak data do not phase up on the
orbital period of 18.4 days. Longer intervals of continuous
observation are required to search for periodicity at longer periods.

Color variations of S\,1063 appear to be wavelength dependent. The $B-V$ color shows little or
no variation as the system brightens. Both $U-V$ and $V-I$ become
bluer by as much as 0.05 mag with increased system brightness.

In contrast, a periodicity comparable to the orbital period was quickly evident in
the S\,1113 data. We show in Fig.~\ref{s1063:1113phot} the $V$ light
curves derived from the 18-day ESO observations and phased to the
radial-velocity orbit solution, which is also shown in
Fig.~\ref{s1063:1113phot}.  The morphology of the light curve is
sinusoidal with a high degree of symmetry. Maximum brightness occurs
at orbital phase 0.5, when the stars are aligned perpendicular to our
line of sight and the primary is approaching. The
system becomes bluer in all bands as it brightens, although again it is most evident in $V-I$.

A time-series analysis on all of the photometric data indicates a
possible period of 2.822$\pm$ 0.001 days. This period is very similar to
the orbital period of 2.823094$\pm$0.000014 days, indicating a close
association. There is some indication that the light curve shape is
not stable. As is evident in Figure 4, three sets of photometric observations
obtained at intervals of 4-6 weeks and 2 years do not repeat to within
the formal photometric errors, suggestive that the source of the
photometric modulation has undergone small changes on these time
scales.

\section{Cluster membership}

We use kinematic evidence to assess the
membership of these binaries in the
M67 cluster. As noted above, Girard et al. (1989) find the 
proper-motion membership probability of S\,1063 to be 97\% and of S\,1113 to
be 98\%. The location of S\,1113 at 3 core radii in projection
makes the argument for its
membership somewhat less secure. However, even weighting by the cluster
surface-density distribution
Girard et al. (1989) give it a membership probability of
94\%, reflecting the fact that S\,1113 lies within the half-mass
radius of the cluster (Mathieu 1983). At the same time, both binaries have
radial velocities within 1  km s$^{-1}$ of the cluster radial velocity.

The proper motions of Girard et al. (1989) are of very high precision, and 
thus their ability to distinguish cluster members and field stars is excellent
(see their Figure 5). Nonetheless, even the highest probability members have
some chance of being field stars. Our radial-velocity survey provides
independent information which permits us to evaluate the percentage of field
stars among the proper-motion high-probability members.
Specifically, we consider the sample of stars with 
proper-motion membership probability greater than 95\%. Within this sample
there are 140 stars with multiple radial-velocity measurements obtained in
the CfA survey and no
indication of velocity variability.
Of these 140 stars, only one has a velocity (25.6 km s$^{-1}$) indicating
non-membership. The remainder show a gaussian velocity 
distribution with a mean velocity of 33.55 km s$^{-1}$ and a 
velocity dispersion of 1.06 km s$^{-1}$. This result remains
true for the subset of 39 stars at projected radii greater than 2 core radii,
which show a mean velocity of 33.45 km s$^{-1}$ and a velocity dispersion of 1.07
 km s$^{-1}$. Based on the single field-star interloper, we
conclude that the field contamination among those high-probability
proper-motion members is of order 1\%.

A fraction of these field-star contaminants may have radial velocities similar to that of the cluster. 
The field radial-velocity dispersion is expected to be 
much larger than the observed cluster velocity dispersion of 0.5 km s$^{-1}$. 
As yet the true field radial-velocity distribution in the
direction and magnitude range of the proper-motion study 
has not been measured. In its place we model the field
radial-velocity distribution with a gaussian having
a typical galactic disk velocity dispersion of 20 
 km s$^{-1}$ (Sparke \& Gallagher 2000) and conservatively center the field velocity
distribution on the M67 radial velocity.\footnote {As a plausibility check, we have examined the distribution of all
radial-velocity non-members in our survey. These stars have a mean radial velocity of 22 km s$^{-1}$
 with an
r.m.s dispersion of 29 km s$^{-1}$. Since the survey selection
was based on proper-motion membership, this distribution
is not independent of the proper-motion distribution and thus may not be appropriate for use in this analysis.
Nonetheless these results give plausibility to the adopted radial-velocity distribution.}
With these assumptions, we find that
4\% of field stars have
velocities within 1 km s$^{-1}$ of the cluster velocity, 
corresponding to two times
the observed velocity dispersion of cluster members. Thus we estimate that
the percentage of field stars with both high proper-motion membership
probability and radial velocities consistent with cluster membership
is no more than 0.04\%.

There are 246 stars in the proper-motion study with membership probabilities
in excess of 95\%. Using the binomial theorem and our estimated probability of 0.04\%,
there is a 0.4\% chance that two of
these stars are field stars with radial velocities consistent with cluster
membership, and a 9\% chance that one such star is a field star. 

Given that the chances of one or even two field stars being
three-dimensional kinematic members of M67 are not entirely negligible, the possibility
that S1063 and S1113 are field stars rather than anomalous members must be explored.
However, as we shall see in the next section, neither of these systems can be completely explained
as field stars. Furthermore, S1063 and S1113 are not the only three-dimensional kinematic members
of M67 without evident explanation; for example, S1072 already has been noted in the Introduction. The probability that all of these seemingly anomalous
cluster members are simply field stars
is indeed negligible. Finally, several of the authors would argue that S1063 and S1113 
have been drawn from the much smaller sample of ROSAT-detected stars, and therefore would conclude that the 
chances that they are field stars are yet smaller than presented here.

\section{Properties of the stars} \label{s1063:prop}
{\bf S\,1063} The most notable property of the orbit of S\,1063 may be
its ordinariness. If S\,1063 fell on the cluster main sequence or subgiant
branch, its orbit would merit no further attention. For example, the
eccentricity of $e = 0.206$ is quite typical for main-sequence and
subgiant binaries in M\,67 with periods $\sim 20$ days (Mathieu, Latham \&\
Griffin 1990).

If the primary star is more massive than the unseen secondary, then
the orbit places an absolute lower limit on the primary and secondary
masses of 0.06 $M_{\odot}$, which is too low to be useful.
Given the anomalous location of S1063 in the M67 CMD, we have no estimate of 
the primary mass based on stellar evolution theory. 
However, we will show below that the primary is an evolved star, with the
surface gravity of a subgiant. Thus we
consider primary masses between 0.7 $M_{\odot}$ and
1.3 $M_{\odot}$, delimited by the main-sequence lifetime equal to the age of the
Galaxy and the upper mass limit on the formation of a subgiant branch.
For this range on the primary mass
the orbit places a lower limit on the secondary mass between 0.23 $M_{\odot}$ 
and 0.34 $M_{\odot}$. 

Evidently the primary star dominates the light at 5187 \AA. As noted
above, the CfA spectra indicate an effective temperature of $T_{\rm
eff} = 5000$ K and a gravity of $\log g = 3.5$. New analyses of the
higher-signal-to-noise spectra of van den Berg et al. (1999) yield
similar measures. Comparison of the V I 6251.83 \AA /Fe I
6252.57 \AA\ line ratio with the spectral analyses of Gray (Gray 1989,
Gray \&\ Johanson 1991) yield a $T_{\rm eff}$ of 5150 K for a dwarf
and 4900 K for a giant. Alternatively, in Fig.~\ref{s1063:class} we
use the spectral diagnostic $I_s$ as defined by Malyuto
\&\ Schmidt-Kaler (1997) to derive a spectral type of G8--K0. Such a
spectral classification indicates effective temperatures similar to
the spectral line results (Schmidt-Kaler 1982, Bessell \&\ Brett 1988,
Bessell et al. 1998).

Detailed analyses of the optical spectra obtained by van den Berg et
al. (1999) indicate that the primary of S\,1063 is neither
a giant nor a dwarf. We have taken two approaches to determining the
surface gravity. First, we have compared the gravity-sensitive Mg b
lines with Kurucz spectra. In Fig.~\ref{s1063:magn} we show the
spectrum of S\,1063 and Kurucz spectra at $T_{\rm eff} = 5000$ K over
a range of $\log g$. Based on the Mg b triplet line shapes, S\,1063 is
neither a giant ($\log g = 2.5-3.0$) nor a dwarf ($\log g =
4.5-5.0$). Spectral fitting to the Mg I b lines (with $T_{\rm eff}$ a free parameter)
results in $T_{\rm eff} = 5000$ K, $\log g$ = 3.5, and $v\sin i$ = 8
km s$^{-1}$, corroborating the results obtained from the lower-signal-to-noise CfA
spectra. However, the referee has correctly noted several disparities of relative line depths in other wavelength regions 
between the observed spectrum and the models, for example comparing the Fe I 5187.9 and Ti II/Ca I 5188.8 lines. 
More detailed spectroscopic analyses and modeling may prove fruitful.

Second, in Fig.~\ref{s1063:class} we compare the values of $I_{s}$
and $I_{2}$ for S\,1063 with those of stars of all luminosity
classes. Again, for stars of similar $I_s$ S\,1063 falls between the
dwarfs and giants. Thus all of the spectroscopic evidence indicates that the
primary of S1063 is a subgiant with an effective temperature
of $T_{\rm eff} \approx  5000$ K. 

Montgomery et al.  (1993) provide UBVI photometry (Table~\ref{s1063:phot}) 
which allows an
analysis of the extinction toward S1063 under the assumption that the
luminosity of the system derives solely from the primary star. 
We initially consider the
extinction of E(B-V) = 0.032 found for M67 by Nissen et al. (1987).
Adopting the extinction curve of Cardelli et al. (1989), the dereddened
colors of S1063 are given in Table~\ref{s1063:phot}. For comparison we also give in
 Table~\ref{s1063:phot}  the colors of a solar-metallicity
4-Gyr 1.37  $M_{\odot}$ subgiant with the
same value of (B-V)$_o$, using the Yale isochrones with the GDK color
transformation (Yi et al. 2001). 
The intrinsic colors of
S1063 are both 0.15 mag bluer in (U-B) and 0.14 mag redder in (V-I).
Furthermore, the effective temperature implied by the (B-V)$_o$ = 1.02 mag is
4577 K, significantly cooler than the effective temperature of 5000 K derived
from spectroscopy.

These color differences can be resolved if the
extinction toward S1063 (or equivalently the
intrinsic color (B-V)$_o$ of S1063) is taken as a
free parameter. For example, we consider the colors of a $\approx$ 5000 K subgiant,
specifically a 4 Gyr 1.35 $M_{\odot}$ subgiant as defined by the Yale
models and presented in Table~\ref{s1063:phot}. 
Such a star implies a reddening of 0.18 mag, and the consequent
intrinsic colors of S1063 are also given
in  Table~\ref{s1063:phot}. The dereddened (U-B)$_o$, 
(B-V)$_o$, and (V-I)$_o$ colors
for S1063 agree very well with the 4989 K subgiant model.

We note that given the observed photometric
variability of S1063 (Section 3.2), detailed modeling of colors as in
Table~\ref{s1063:phot} must be considered with care. S1063
shows little variation in (B-V), and so combined with the 
spectroscopic derivation of the effective temperature 
the  argument for enhanced extinction toward S1063 remains unchanged.
The argument is also not
sensitive to the surface gravity of the primary star, with which the colors
vary only slightly. On the other hand,
the observed (U-B) and (V-I) of S1063 
show variations of up to 0.05 mag peak-to-peak, and
so there is uncertainty in the use of the single-epoch Mongtomery et al.
photometry to specify the nature of the system. (Perhaps surprisingly, the
(U-B) measurements of Mongtomery et al., Racine (1971) and Sanders (1989)
agree to within 0.01 mag; the (B-V) measurements agree to within 0.02 mag.)

The amount of inferred extinction also depends on the choice of stellar
model. The Yale models have also been transformed to observables using the
LCD transformations (Yi et al. 2001).
In Table~\ref{s1063:phot} we repeat the subgiant analysis with
these colors. While the derived reddening of E (B-V) = 0.14 mag 
is somewhat smaller than with the GDK transformations, it
nonetheless remains larger than the extinction toward M67 itself.

Interestingly, correcting for a reddening of E (B-V) = 0.18 mag moves S1063 
to a position 0.15 mag in V below the base of the M67 giant branch. Given an observed
variation in V of 0.15 mag, S1063 remains a candidate for a normal cluster
subgiant if an explanation for the higher extinction can be found. Adopting
this reddening, the extinction curve of Cardelli et al. (1989), the bolometric correction of the Yale model, 
and a true distance modulus to M67 of 9.6, we can derive a radius of 2.4 $R_{\odot}$ and a
luminosity of 3.3  $L_{\odot}$ for the primary star
(Table~\ref{s1063:phot}). Given a log g = 3.5 this radius implies a mass of 0.7 $M_{\odot}$; however
we have not determined the gravity to better than a factor 2-3.
We stress that all of these values are only meaningful given cluster membership.

Alternatively, the colors of S1063 can be modeled with a background
giant star having an
effective temperature somewhat in excess of 4900 K.
For specificity, we show in  Table~\ref{s1063:phot} the intrinsic colors 
of a 0.8 Gyr 2.3  $M_{\odot}$ giant, using the Yale models with the LCD
transformation. This model agrees well with the intrinsic colors of S1063
given E (B-V) = 0.14 mag. 
With  $M_v = 1.1$ mag, the distance of such a giant would be 2.8 kpc,
roughly 1.5 kpc above the galactic plane. 
Placing S1063 at such a distance does not encounter significant difficulties with
other observations. For example, the resulting luminosity in the Ca H and K lines
is typical for RS CVn's with giant primaries (van den Berg et al. 1999). Similarly,
the X-ray luminosity would become 8 x 10$^{31}$ ergs/sec (0.1 - 2.4 keV). Examination of Figure 4
in Belloni et al. (1998) shows such a flux to be at the
  high end of observed X-ray
fluxes from similar RS CVn primary stars.

However, the large distance does {\it not} explain the required extinction. 
The extinction
maps of Schlegel et al. (1998) indicate a Galactic extinction in the direction of M67
of E(B-V) = 0.03 mag, essentially placing all of the Galactic extinction in front of the
cluster. This giant model for the primary also has the modest difficulty of a lower
surface gravity (log g = 2.9) compared with the
log g = 3.5 derived from the spectral analyses. 

In summary, these color analyses suggest a reddening of S1063 of order 0.15 mag more
than can be attributed to Galactic extinction in the direction of M67. The system might be interpreted as a
cluster subgiant or as a distant ($\approx$ 3 kpc) halo giant, but neither
interpretation naturally explains the enhanced reddening. 
A conjecture is that the reddening might be due to circumstellar or 
circumbinary material local to the system. 
Indeed, the
slow but substantial V brightness
variation observed in this system might be attributed to variations in local
extinction; however such an explanation is confounded by the lack of 
associated variation
in (B-V). 
We also note that the system
shows no evidence of a near-infrared excess. Infrared colors have been
obtained from the 2MASS survey and are given in Table 5. Theoretical
colors are also available from the Yale models with the LCD
transformation, also given in Table 5 for the 1.35  $M_{\odot}$ model
at 4 Gyr. 
The dereddened observed colors and theoretical colors
agree to within the 4\% measurement errors on the colors. 
Reducing these errors and extending measurements to longer wavelengths
would be
worthwhile.

Of course, these
color analyses presume that the colors of the primary star
are not abnormal for a 5000 K star, and that the
composite light of the system derives only from the primary photosphere.
We encourage more careful spectroscopic analyses of S1063, among other things
to precisely compare abundances with cluster members,
to better determine the surface gravity, and to search for evidence of
extincting matter. We also recommend
photometric observations over a broader wavelength range
to search for other contributors to the
composite light.

Finally, we briefly discuss the rotation of the primary star
in the context of an M67 1.3 M$_{\odot}$ subgiant with
a system E (B-V) = 0.18 mag.
The non-detection of the secondary in our spectra
requires that the secondary star have $V>16$, roughly two magnitudes
fainter than the primary (here taken as equivalent to the composite
light). If the secondary is a main-sequence star in the cluster, 
then based on this
magnitude limit and extinction
the Yale isochrones give an upper limit on the mass of
the secondary of 0.9  $M_{\odot}$ and an associated lower limit on the
inclination angle of 28 deg from the orbital solution.
For aligned rotational and orbital axes and a measured vsin i of 6 km s$^{-1}$, 
the true equatorial rotation velocity could be as large as 13 km s$^{-1}$. 
For a primary radius of 2.4 $R_{\odot}$ the rotational period could be as 
short as 9.5 days. Such a
short rotation period would be consistent with the observed X-ray
emission, chromospheric activity, and possible flares. Perhaps most
plausibly the pseudo-synchronous rotation period of 14.64 days (cf. Claret
\& Giminez 1993), requiring an inclination angle of $i = 46^{\circ}$, 
is permitted by the primary mass range and the upper limit on the secondary
mass.
However, we remind the reader that the
photometric data show no evidence of periodicity at this
rotation period, nor is there strong H$\alpha$ emission from the
primary (van den Berg et al. 1999).

{\bf S\,1113} The spectroscopic detection of the secondary permits
substantially more information to be derived for this binary. Here we
analyze the system first from a distance-independent
geometric perspective, and then we take a distance-dependent
photometric approach.

For the geometric argument, we presume that the stellar rotational periods are
synchronized with the orbital period, and that the rotational and orbital
axes are aligned. Both are expected for a circularized orbit, since alignment
and synchronization happen on shorter timescales than circularization 
(Hut 1981), and synchronization is further supported by the observed
photometric variation on a period very similar to the orbital period.

Given these two assumptions, 
the mean density of either star can be written as
\begin{equation}
\rho \propto { {\left(M \sin ^3 i \right)} \over {\left(P \: v \sin i \right)^3} } \, .
\end{equation}
The mean density is thus independent of the unknown inclination angle and 
completely determined by the observables Msin$^3 i$, P, and vsin i. Using
the values for these observables and T$_{eff}$ given in Section 3.1, both stars
can be placed on the density-effective temperature diagram, as shown in 
Figure \ref{fig-rho-temp}. 
Also shown in Figure \ref{fig-rho-temp} are solar-metallicity Yale isochrones for a set of
ages. The 4 Gyr isochrone, comparable to the age of M67, is shown as a dashed 
line.

The mean density of the  secondary indicates a main-sequence star 
with an age of between 1-10 Gyr. 
While the uncertainties are too large to precisely determine the secondary age,
the inferred secondary mass varies little as a function of age. For a 4 Gyr isochrone
the mass of the secondary star is 0.93 $M_{\odot}$, and the range on mass
implied by the uncertainties extends from 0.89 $M_{\odot}$ to 0.97 $M_{\odot}$. For a mass of 0.93 $M_{\odot}$, 
the 
dynamical mass ratio implies a mass
for the primary star of 1.33 $M_{\odot}$ with a range of 1.27
$M_{\odot}$ to 1.38 
$M_{\odot}$. The inclination of the system becomes
48$^{\circ}$, and the semi-major axis is 11.0 $R_{\odot}$.
The primary and secondary
radii, presuming synchronous rotation for both stars, 
are 4.0 $R_{\odot}$ and 0.90 $R_{\odot}$, respectively. 

This derived mass for 
the primary star is notable in that the mass of a 4 Gyr giant 
is 1.37 $M_{\odot}$. That is, the derived 
mass of
the primary star is very similar to that expected for an evolved M67 member.
Similarly, the location of the
primary star in Figure \ref{fig-rho-temp} is closely consistent with the 4 Gyr isochrone of M67,
although the uncertainties in effective temperature permit a wide range of
ages for the primary star.

A second independent analysis of the primary and secondary stars can be done
from the observed photometry and the spectroscopically determined 
magnitude difference of
$\Delta V=1.23$ mag, assuming no other contributors to the composite light. 
Adopting again the photometry  
of Montgomery et al. (1993) for the composite system, the primary and
secondary $V$ magnitudes are derived immediately from the measured flux
ratio (Table 6). 

To obtain colors for the primary and secondary stars, we begin by presuming
membership in M67 and use the 5 Gyr Yale models
(GDK transformation). The derived secondary $V$ magnitude, a reddening of
E (B-V) = 0.032 mag, and an apparent
distance modulus of 9.6 mag (Montgomery et al. 1993) gives a secondary
absolute magnitude $M_v$ = 5.6, indicating a
4 Gyr 0.90
$M_{\odot}$ main-sequence model for the secondary star 
(Table~\ref{s1063:1113decomp}). 
Notably, this mass is similar to the secondary mass of 
0.93 $M_{\odot}$ derived from independent geometric arguments. It is also
encouraging that the
effective temperature of the 0.90 $M_{\odot}$ cluster member is 5311 K, 
within the measurement uncertainties of 
the spectroscopically determined effective temperature of 5500 K (Section 3.1).
\footnote {Turning this argument around, we adopt  a 4 Gyr, 5500 K star as
a model for the secondary.
The implied mass is 0.95 $M_{\odot}$ with an absolute V magnitude of M$_v$ = 5.3.
Unreddened, the distance modulus to such a star is 10.0 mag; given reddening
this is an upper limit.
Thus, if the secondary is a main-sequence
star, S1113 cannot be located far beyond M67.}

If we adopt the model colors for a 0.90 $M_{\odot}$ 5 Gyr main-sequence star, 
then we can 
determine colors for the primary star as
presented in Table~\ref{s1063:1113decomp}.  These
results for the primary and secondary are also plotted as open boxes in
Fig.~\ref{s1063:cmd}. 
The primary lies somewhat further
below the subgiant branch and further to the red 
than indicated by the composite light.

This color deconvolution leaves an interesting loose end, in that the
primary shows a (U-B) blue excess of 0.3 mag relative to either main-sequence
or giant colors. 
A giant model selected to match the dereddened primary (B-V) color is shown in Table~\ref{s1063:1113decomp};
the (U-B) color of this model is significantly redder than derived for the primary star. The (U-B)
color of a main-sequence star of the same (B-V) is 0.89.
Both the U-band measurements of Sanders (1989) and
Montgomery et al. (1993) support such an excess, although their measurements
differ by 0.07 mag and possibly indicate variability.
This excess ultraviolet light may in fact be associated 
with the primary star, or it may derive from elsewhere in the system. 

Since the gravity of the primary star is weakly constrained by the spectroscopy, 
we use both main-sequence
and giant models of the same B-V color as the primary to provide bolometric corrections
and luminosities at the distance of M67. The giant model implies
a luminosity for the primary of 2.0 $L_{\odot}$ and an effective
temperature of 4534 K. The  main-sequence model provides
a luminosity of 2.1 $L_{\odot}$ and effective temperature
of 4726 K. Both of these temperatures are
somewhat lower than the spectroscopically determined effective 
temperature of 4800 K (Section 3.1).

Unlike the secondary star,
the model for the primary star inferred from these photometric arguments 
is markedly different from the model for the
primary developed from the geometrical arguments. The 
photometric radius for the
primary of 2.0 $R_{\odot}$ is significantly smaller than
the geometric radius of 4.0 $R_{\odot}$, or equivalently the photometric
luminosity of 2.0 $L_{\odot}$ is substantially smaller than the geometric
luminosity of 7.7 $L_{\odot}$.

This discrepancy between the primary luminosities derived from geometric and
photometric arguments can be cast in a distance-independent manner. The
geometric argument provides a distance-independent prediction for the ratio of
primary to secondary V flux. Continuing to assume synchronization and alignment
of axes, the ratio of primary to secondary radii is the same as the ratio
of vsin i. Using also the ratio of our spectroscopically determined
effective temperatures
(Section 3.1), the ratio of bolometric luminosities then becomes 
$L_{2,bol}/L_{1,bol} = 0.085$. 
Using the Yale bolometric corrections, we find that 
$L_{2,V}/L_{1,V} = 0.11$. 
This ratio is
significantly smaller than the measured flux ratio at V of 0.352 $\pm$ 0.016
(Section 3.1). The sense is consistent with the primary luminosity 
being a factor 3 less 
luminous than predicted by the geometric argument, as in the last paragraph.

\section{Discussion} \label{discussion}
{\bf S\,1063} 
The binary S1063 is remarkable in three ways:
low apparent brightness compared to M67 subgiants of similar color, enhanced reddening compared
to Galactic extinction along the line of sight, and
weak H$\alpha$ emission observed at a velocity different from
the primary star velocity. In addition, the long
timescale light variations are not easily explained if they are
aperiodic. If they are periodic on timescales longer than 18 days, 
then they likely are due to star spots 
on the primary star and indicate that the primary is 
not rotating pseudosynchronously.

Given both its close kinematic association with M67 and projection
upon the core of the cluster, we have explored explanations for S1063 in the 
context of cluster membership. 
In this context, we have previously inferred
a radius of 2.4 $R_{\odot}$ for the primary
star. As such, the
primary star does not approach its critical surface at
periastron passage. Thus the radius of the star is not confined by the
presence of the secondary, nor is there motivation for present mass
transfer. Similarly,
the eccentricity of the orbit does not suggest either a large evolved star 
or mass transfer in the past, 
both of which would likely have circularized
the orbit. 

We have explored the possibility that rather than being an underluminous
subgiant the primary star is instead an overluminous main-sequence star. 
For example, a recent merger of two main-sequence stars
would deposit kinetic energy into the merging stars.
Here we consider two types of mergers: coalescence and collision. In the
coalescence scenario, the S\,1063 system was originally a triple
system in which the present secondary was the tertiary star. If we
assume that the orbital period that we observe now was equal to the
period of the original tertiary star, we can use stability arguments
to place an upper limit on the period of the original inner binary
star. Using the coplanar formulation of Mardling \&\ Aarseth (1998)
for the stability of a triple system, and adopting a mass ratio of
$q = 0.5$ (outer star to inner binary), we find that an original inner
binary would have had an orbital period of less than 2.22 days. Such a
period is physically permitted, and so that S\,1063 was formerly a
triple system is possible.  Coalescence via multiple mechanisms is
plausible on time scales of a few Gyr (St{\c e}pie\'n 1995). Four
contact binaries are known in M67 (see light curves in van den Berg et al.
(2002) and Stassun et al. (2002)), and the
old (6 Gyr) cluster NGC 188 has many contact binaries (Baliunas \&\
Guinan 1985, Rucinski 1998).

A collision scenario has been suggested by Leonard \&\ Linnell
(1992) as a possible mechanism for
the formation of some blue stragglers in M\,67. 
In the low-density environment of an open cluster, collisions
of single stars are improbable. However, binary-binary encounters have
substantially higher probabilities, and in the course of such
encounters the probability of a stellar collision is not
negligible. In this scenario, the present S\,1063 would consist of
three of the stars involved in the binary-binary encounter, two having
merged into the primary and a third being the present secondary. An eccentric
orbit is a natural result of such a resonant encounter.
A prediction of either of these merger scenarios is that the
resulting star will rotate rapidly. In the coalescence scenario rapid
rotation is expected as a result of synchronous rotation prior to the
merger. With respect to the collision scenario, the simulations of
Sills et al. (2001) show that a collision product is born with a high
rotation rate. We argued in Sect. 5 that for aligned
rotation and orbital axes, the true equatorial rotation velocity
could be as large as 13 km s$^{-1}$, and for a primary radius of 2.4
$R_{\odot}$ the rotational period could be as short as 10 days. However
such a rotation period is still long compared to those expected from
mergers or collisions. In the collision scenario, of course, the
primary rotation axis and the orbital axis need not be aligned,
particularly since tidal circularization has not come to completion.
As such, the surface rotation velocity of the primary could be
significantly higher. A high inclination angle could also possibly
explain a lack of periodic photometric modulation.

Most problematic, these impulsive origins must face the challenge that
-- even if the merger product would briefly have the properties of the
S\,1063 primary star -- the thermal time scale of the primary is very
short compared to the cluster lifetime. If, for the sake of example,
we adopt a primary mass of 0.7 $M_{\odot}$, the present thermal time scale
$E_{\rm pot}$/$L$ is only 3.4 Myr. A merger product would be expected to
readjust to its new mass on such time scales (and perhaps become a present
or future blue straggler depending on the combined mass).
Given such short
time scales, the probability of observing the primary prior to its new
equilibrium state is very low. On the other hand, should a star be found
shortly after a merger, the presence of residual 
circumstellar material resulting
from the merger might also be expected, and represents a possible
explanation for the enhanced reddening of this system. Lastly, if formed recently
such a system may not yet be pseudo-synchronized.

We stress that the nature of the secondary is unknown;
we know only that it is substantially fainter than the primary 
at $V$. Thus the secondary is permitted to
be a white dwarf. An isolated white dwarf of the cluster age
would have a mass of order 0.6--0.7 $M_{\odot}$
(Wood 1992, Richer et al. 1998); a white dwarf deriving from a
prior mass transfer scenario could be much different. 
A hot source of radiation might help explain
 the weak H$\alpha$ emission observed at a velocity different from
the primary star velocity.
How the secondary could become
a white dwarf without circularizing the orbit during its prior
post--main-sequence evolution would need to be explained; again,
a dynamical exchange might play a role.


{\bf S\,1113} In Section 5 we derived  the properties of the stars in S1113 
from two independent lines of reasoning,
geometric and photometric. The two arguments produced quite 
different conclusions regarding the
luminosity ratio of the stars, or similarly regarding the nature of the 
primary star. To bring these two lines of reasoning into
agreement, we must give up a basic premise of at least one of the arguments.
Here we explore two possibilities, and then discuss the measurement uncertainties.

a) Both the geometric and photometric arguments converge on a 0.9 $M_{\odot}$ 
main-sequence secondary. 
This suggests that the resolution of the contradiction
in luminosity ratios might be found in
the primary star. Under the assumption of membership in M67 the geometric
argument must be found wrong, for its implied 4800 K primary star of radius 
4.0 $R_{\odot}$ would be 1.2 mag more luminous in V than observed. The problem
cannot be solved with extinction, since the reddening vector for the S1113
primary does not pass through an M67 4.0 $R_{\odot}$ giant. 

Examining the premises of the geometric argument, we first note that
relaxing the assumption of alignment of orbital and rotational axes  cannot
in itself resolve the problem. Even if the inclination angle of the primary
rotation axis
is taken to be 90$^{\circ}$, its radius is reduced only to 3.0 $R_{\odot}$. 
(Recall that the primary radius derived from the photometric argument is
2.0 $R_{\odot}$.)

Supersynchronous rotation of the primary leads to more success.
Specifically, a rotation period of 1.4 days - or supersynchronous rotation by a
factor of 2 - would bring the primary
radii derived from both the geometric and photometric arguments into agreement 
on a primary radius of 2.0  $R_{\odot}$.

Such supersynchronism is not expected theoretically given the close proximity of the
stars, the primary's large radius, and the surface convection zones on both stars.
Unless continuously driven, the duration
of a supersynchronous state would be short. Furthermore,
supersynchronous rotation of the primary would also require that the observed variability
of the composite light with the orbital period must derive from elsewhere in the
system. A cool spot on the secondary seems an unlikely origin. Given a secondary/primary
V luminosity ratio of 0.35, a 38\% light modulation from the secondary would be
required to reproduce the observed 10\% variation in the composite light at V. 

We note that making the secondary subsynchronous would imply a larger, more
luminous secondary which would also resolve the flux ratio discrepancy.
However the consequent change in mean density moves the secondary off the
main sequence into a domain not occupied by standard stellar evolution (Figure 7).
While perhaps this is the case, this path permits essentially unconstrained modeling
of the binary.

Finally, the unusual 
CMD position of S1113 is not easily solved by enhanced extinction of a
cluster member. Large extinctions (in excess of 1 mag)
along a standard reddening vector 
would imply a primary star near the top
of the M67 main sequence, with much higher effective temperatures than
found spectroscopically.

b) Alternatively, we consider arbitrarily discounting the 
spectroscopically determined flux ratio at V of
0.32 (Sec. 3.1)
and adopt the geometric model for the system. This model indicates
a secondary/primary flux ratio at V of 0.11. Corresponding decomposition
of the composite V magnitude leads to a primary V = 13.88 and a
secondary V = 16.31. The geometric model indicates a 0.9 $M_{\odot}$ main-sequence
secondary, and so without considering additional reddening the secondary
V magnitude places the binary 1 mag beyond M67. A similar conclusion is reached by comparing the luminosity of a 4800 K, 4.0 $R_{\odot}$ primary star to the
composite light. 
At this
distance the 1.3 $M_{\odot}$ primary becomes very similar to a star at the base of
the M67 giant branch, and as such can be explained by standard stellar evolution
theory (albeit not as a cluster member).

The larger primary radius of 4.0  $R_{\odot}$ implied by the
geometric model represents a large fraction of the primary Roche radius.
The Roche radii are 4.5 $R_{\odot}$ and 3.8 $R_{\odot}$ for the
primary and secondary, respectively. An equipotential surface  about the primary
whose volume equals that of a 4.0 $R_{\odot}$ sphere extends 74\% of the
way to the $L_{1}$ point. 
Thus, within the geometric model, the role of mass transfer in the evolution of
S1113 merits consideration. Evidently the 0.9 $R_{\odot}$ main-sequence
secondary lies well within its Roche radius.

Interestingly, both the phasing of the
periodic photometric variation and the ultraviolet excess are suggestive of a hot
spot near the secondary star powered by an accretion stream. Such spots are formed on the
following side of the secondary (e.g., Flannery 1975), in qualitative agreement with
the phasing of the light and velocity curves shown in Figure 4.
The 10\% system brightness variation at V 
requires a mass accretion rate of 2.5 x 10$^{-8}$  $M_{\odot}$/yr, presuming that
the energy is released by a 10,000 K spot on the surface of the secondary star.

We note that at present no spectral signatures of such a hot spot have been seen.
Van den Berg et al. (1999) did find broad H$\alpha$ 
from S1113, but this emission was kinematically
associated with the primary star. They argued that this emission could be consistent
with the higher chromospheric activity driven by the rapid rotation of the star,
in analogy to V711 Tau.
In this context, we note that the photometric variation can also be well
modeled by a large cool spot on the primary star. For example, we have
reproduced the light variations with the addition of a cool spot in the
upper hemisphere of the primary and located 90$^{\circ}$ from the
major axis of the orbit. Specifically the spot properties are: latitude
40$^{\circ}$, longitude 270$^{\circ}$, radius 27$^{\circ}$,
temperature 0.82 of the stellar surface temperature. The location of
the spot at 270$^{\circ}$ longitude is motivated by the observations
rather than an independent physical argument.

The large extent to which the primary star fills its Roche lobe in this
model implies
that its shape is significantly asymmetric.
Using the Wilson-Devinney formalism we have
investigated the expected magnitude of photometric variations due to the
anticipated ellipsoidal shape of the primary. Adopting the spot parameters above,
a temperature of 4800 K for the primary star, a limb darkening
coefficient (linear law) of 0.6 for each star, a gravity-brightening
coefficient of 0.32, and a reflection coefficient of 0.5, we find the
peak-to-peak ellipsoidal variations to be 0.06 mag in $V$. Most
importantly, these ellipsoidal variations produce a double-peaked
light curve over the orbital period, in marked contrast to the
observed single-peaked light curve. It is possible that these
small ellipsoidal variations have gone undetected in contrast to the
larger photometric variations of the system. For example, in 
Fig.~\ref{s1063:ellips} we show the light curve combining both
the cool spot described above and the ellipsoidal variations. Evidently the
ellipsoidal variations are lost to inspection in this synthetic light
curve, which in fact looks very similar to the observed light curve.

To summarize, the geometric line of reasoning suggests that S1113 is a field
binary located behind M67. In this scenario it may be a detached RS CVn whose 
rapid rotation is producing the several emission diagnostics of enhanced
chromospheric activity as well as a large spot. Alternatively,
mass transfer may be underway, producing a hot spot in the vicinity of
the secondary star.
However, to adopt this interpretation of the system,
both the kinematic association with M67 and the agreement
of the primary mass with the M67 turnoff mass must be taken as merely coincidental.
In addition, the measured V flux ratio must be ignored. As we discuss below, we
find this last requirement to be a serious counterargument.

c) Measurement uncertainties
We have explored whether the discrepancy between the conclusions of the
geometric and photometric analyses are the result of our measurement uncertainties.
On the photometric side, broadband photometric
measurements of S\,1113 are several and corroborative. 
The effective 
temperature for the primary derived from photometric colors 
is similar to the
effective temperature derived from our
high-resolution spectra, indicating that both determinations of the effective
temperature are reasonable. 
Finally, we find it unlikely that the bolometric correction for the primary
star
overestimates the luminosity by a factor 3, particularly given the observed
excess flux in the U-band. However, given that only UBV photometry is available,
photometry over a wider range of wavelength is much needed.

The geometric luminosity ratio can be brought into accord with the observed V
ratio if the effective temperatures and projected rotation velocities of both
stars are all adjusted by 1.5 $\sigma$ in the appropriate senses. While {\it ad hoc}
and improbable, the required adjustments are small enough relative to their uncertainties
that further precise measurements of these quantities are in order.

The measured flux ratio at V is pivotal in this discussion.
External checks of this quantity on other eclipsing binary
systems, using  similar spectroscopic
material and with the same techniques used here, have not shown
significant systematic errors.  For example, a comparison with
independent determinations available from the light-curve solutions of
double-lined eclipsing binaries typically agree within 10\% or better
with our spectroscopic determinations (see, e.g., Lacy et al.\ 1997;
2000).  Thus we have no reason to believe that our flux-ratio 
measurement for S1113 would be in error by as much as a factor 3. Nonetheless, given the importance of this 
measurement we encourage additional observational study.

In closing, we note that Hurley et al. (2001) have suggested that S1113 is subluminous due to an evolutionary 
response to mass transfer. Albrow et al. (2001) identify stars below the subgiant branch of the globular cluster 47 Tuc 
which they
note are similar to S1063 and S1113. They suggest that such systems may result
from mass transfer, with the subsequent contraction of the primary Roche radius
deflating the primary star and making it less luminous. Our analyses indicate that, considered as a cluster member (i.e., 
the photometric model), both stars in S1113 presently fall well within their Roche radii. Put another way, if the 
primary of S1113 fills its Roche lobe (e.g., the geometric model) and has an effective temperature of 4800K, the 
observed brightness places it behind M67 (and as noted above this conclusion
of non-membership cannot be removed via extinction arguments). 
Thus at least presently 
S1113 would not seem to be a candidate for a Roche-filling cluster binary.

\section{Summary}
\begin{figure}[ht]
\centerline{\includegraphics[width=15cm]{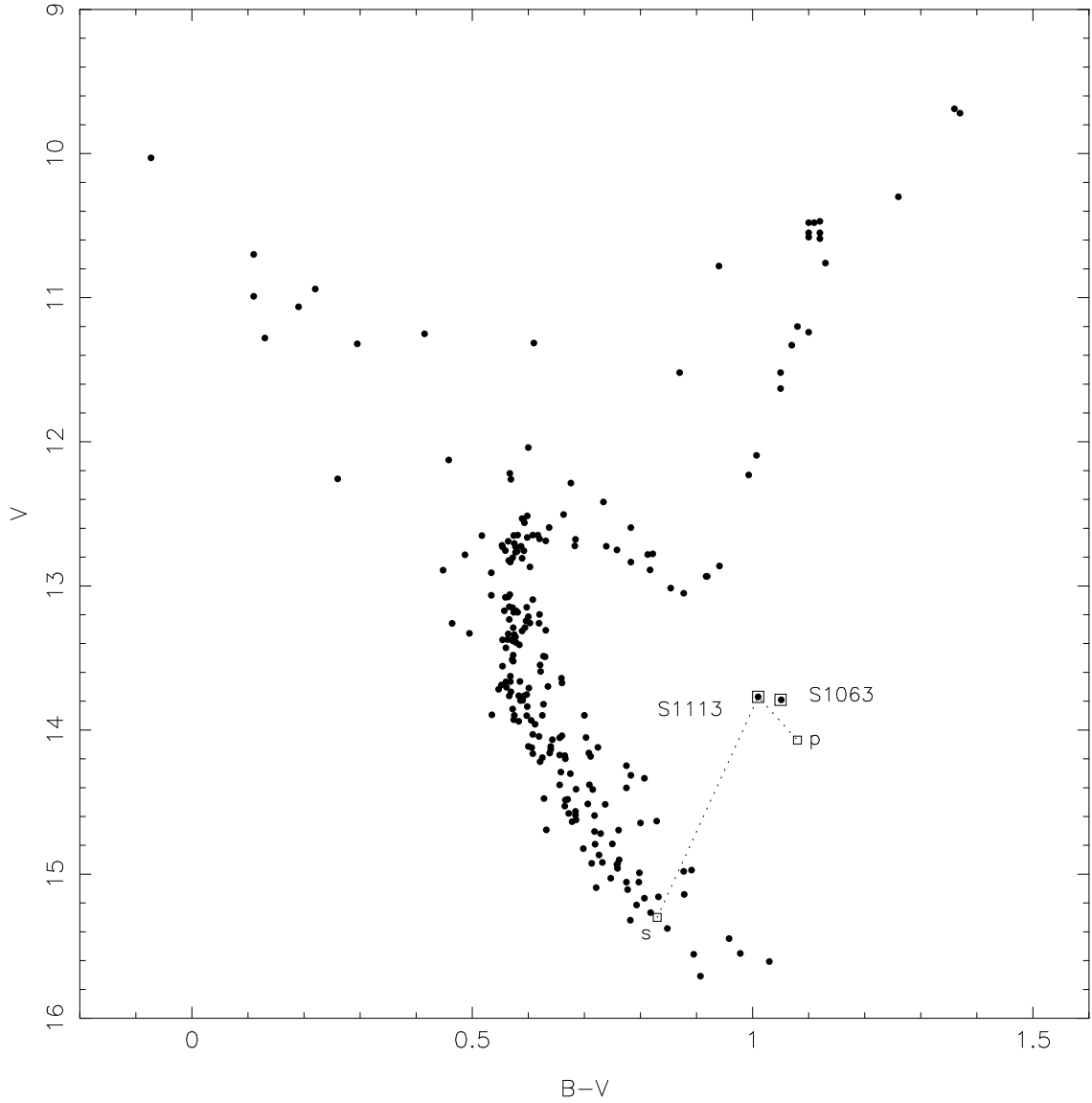}}
\caption{{\small Color-magnitude diagram of M\,67 based on the
photometry of Montgomery et al.  (1993) and including only stars with
proper-motion membership greater than 80\% (Girard et al.  1989). The
suggested sub-subgiants -- S\,1063 and S\,1113 -- are highlighted.  We
also show the decomposition of the light of S\,1113 into the primary
(p) and secondary (s) which applies if the secondary is a
main-sequence cluster member (see also Sect.~5).}}
\label{s1063:cmd}
\end{figure}

\begin{figure}[ht]
\centerline{\includegraphics[width=10cm,bb=73 244 543 612]{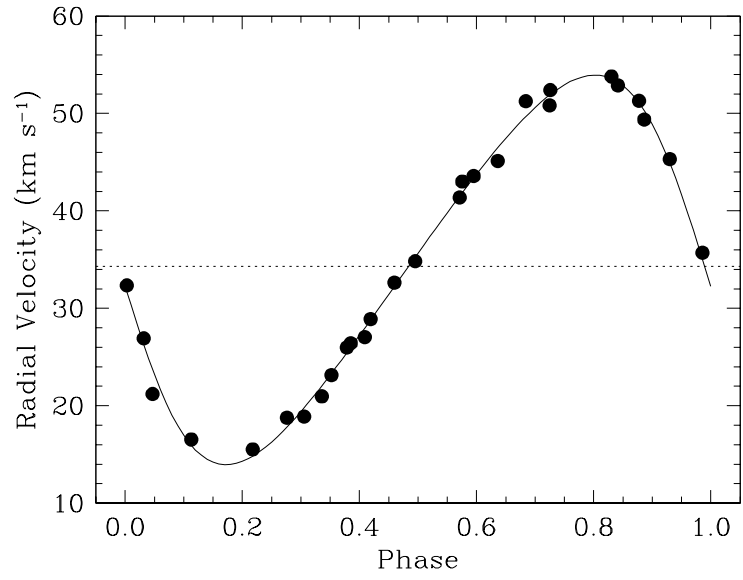}}
\caption{{\small The theoretical velocity curve for S\,1063 shown with the phased
radial-velocity measurements. The dashed line shows the center-of-mass
velocity of the binary.}}
\label{s1063:rv1063f}
\end{figure}

\begin{figure}[ht]
\centerline{\includegraphics[width=10cm]{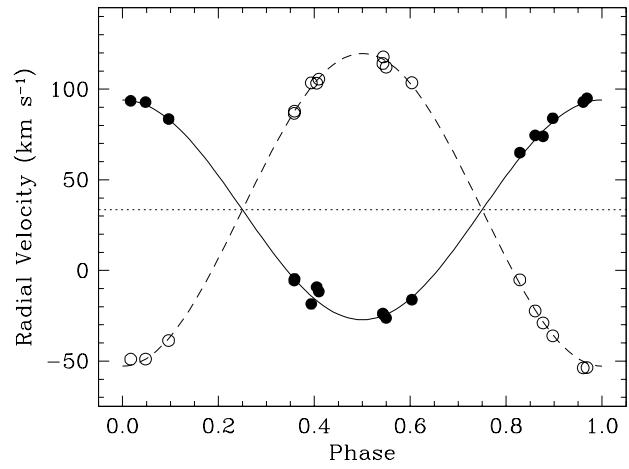}}
\caption{{\small The theoretical velocity curves of S\,1113 shown with the phased
radial-velocity measurements for the primary (filled circle) and
secondary (open circle) stars. The dashed line shows the
center-of-mass velocity of the binary.}}
\label{s1063:rv1113f}
\end{figure}

\begin{figure}[ht]
\centerline{\includegraphics[width=12cm]{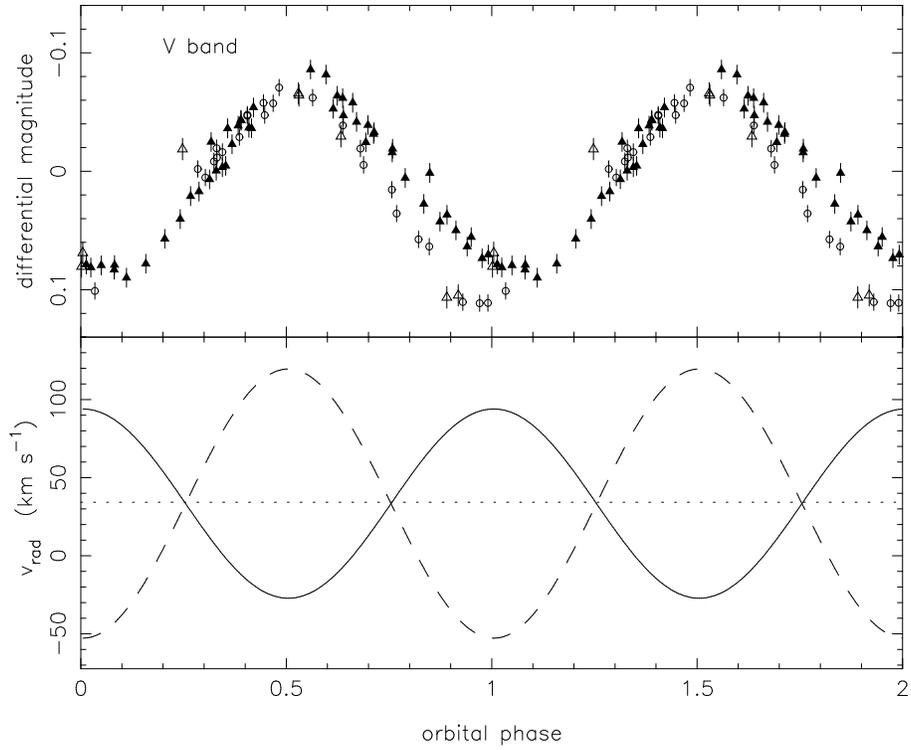}}
\caption{{\small Light curve (top) and primary and secondary orbit curves
(bottom) for S\,1113 phased on the orbital period of 2.823094 days. The
light curve shows $V$ data taken in January 1998 (open circles),
February and March 1998 (filled triangles) and February 2000 (open
triangles), taken from van den Berg et al. (2002). Notice
that maximum light brightness occurs when the primary star is moving
toward us (in projection).}}
\label{s1063:1113phot}
\end{figure}

\begin{figure}[ht]
\centerline{\includegraphics[width=14cm]{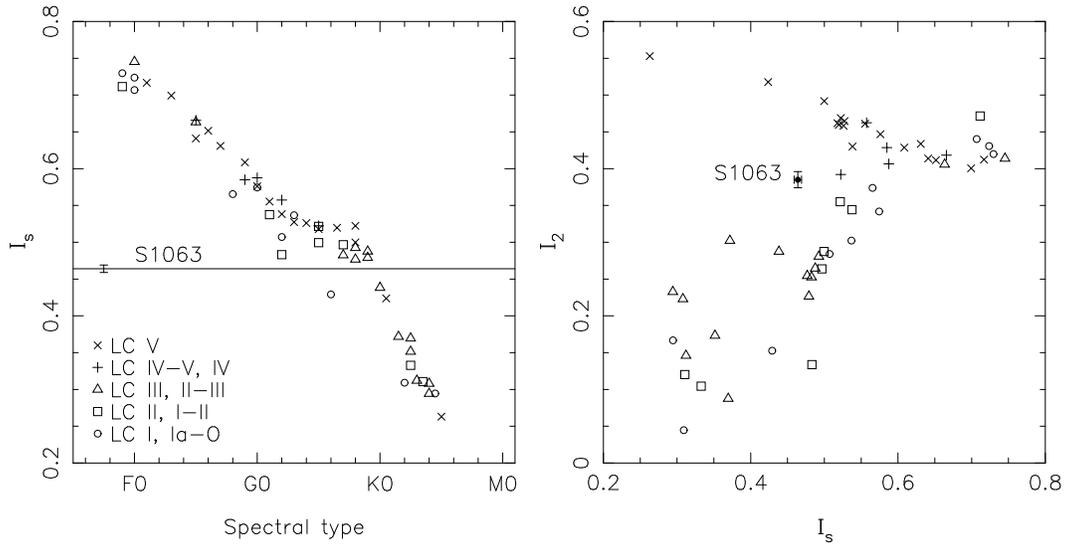}}
\caption{{\small Spectral classification of S\,1063 according to the
classification criteria of Malyuto \&\ Schmidt-Kaler (1997). Left:
$I_S$, based on features between 4215 \AA\ and 4360 \AA\, and 5125
\AA\ and 5290 \AA, is an index for spectral type and places S\,1063
(solid line) between type G8 and K0.  Right: $I_2$, based on features
between 4120 \AA\ and 4280 \AA\, is an index for luminosity class and
places S\,1063 between dwarfs and giants.}} \label{s1063:class}
\end{figure}

\begin{figure}[ht]
\centerline{\includegraphics[width=12cm]{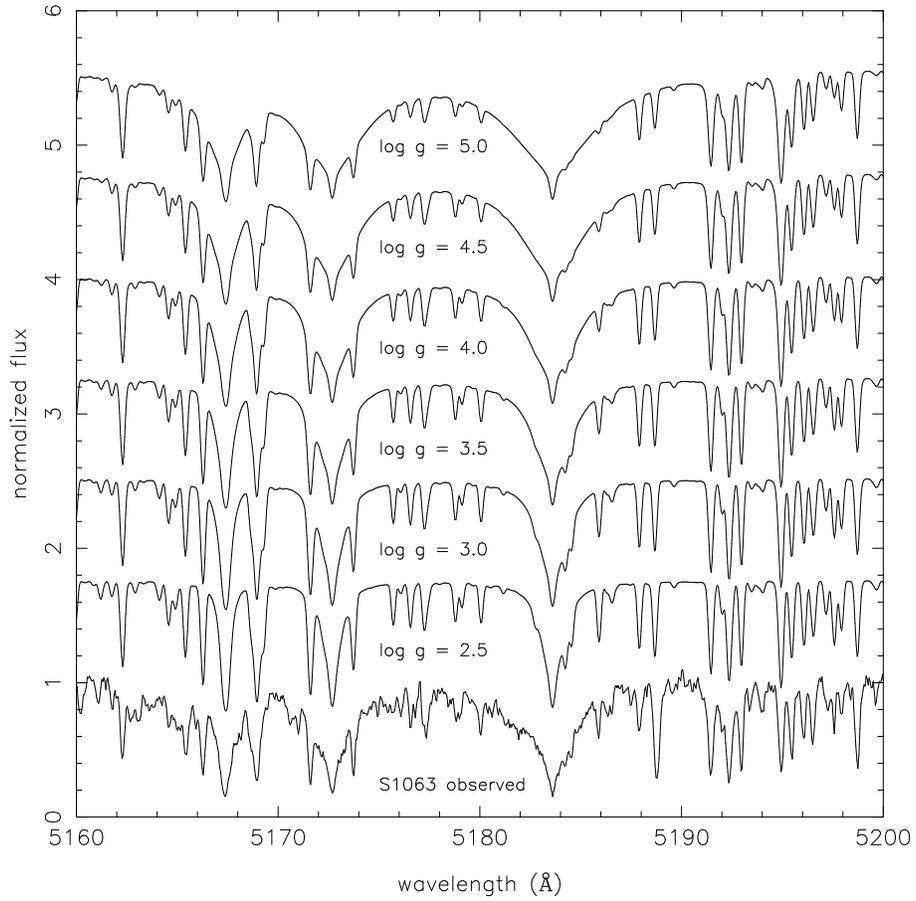}}
\caption{{\small Comparison of S\,1063 spectrum with synthetic
spectra of effective temperature $T_{\rm eff}$ = 5000 K, $v\sin i$ =
8 km s$^{-1}$, and differing surface gravities (Kurucz 1979). The
spectra are normalised to the continuum flux and shifted vertically
in steps of 0.75 flux units. S\,1063 is best fit by the synthetic
spectrum with $\log g$ = 3.5.}} \label{s1063:magn} \end{figure}

\begin{figure}[ht]
\centerline{\includegraphics[width=12cm]{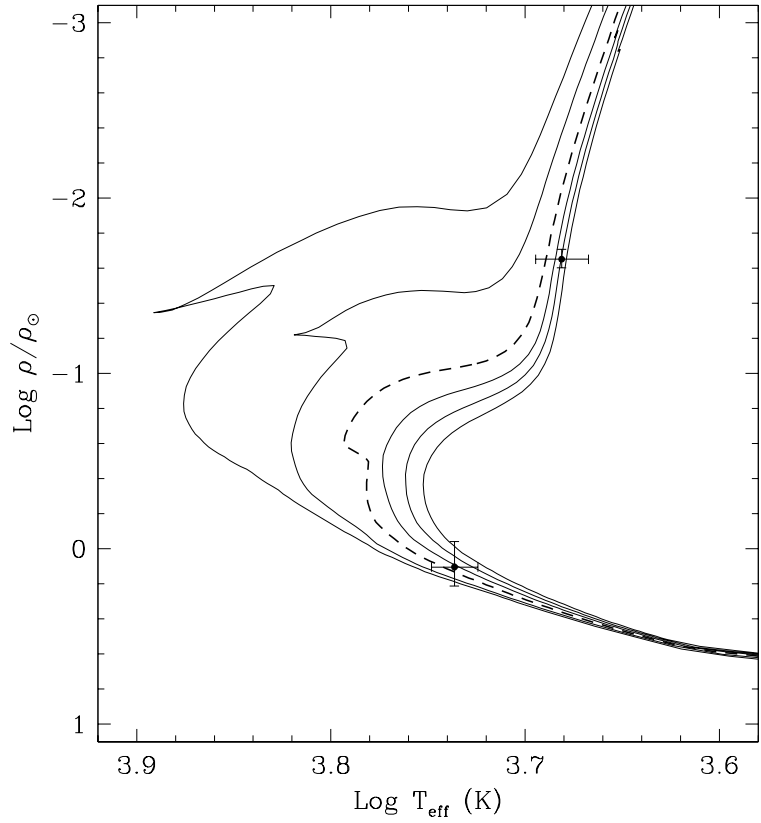}}
\caption{\small Density - effective temperature diagram. The lower density
datum represents the primary star, the other datum is the secondary
star. The curves in the figure are solar-metallicity Yale isochrones for ages
1, 2, 4, 6, 8 and 10 Gyr.
The 4 Gyr isochrone, comparable to the age of M67, is shown as a dashed 
line. }
\label{fig-rho-temp}
\end{figure}

\begin{figure}[ht]
\centerline{\includegraphics[width=12cm]{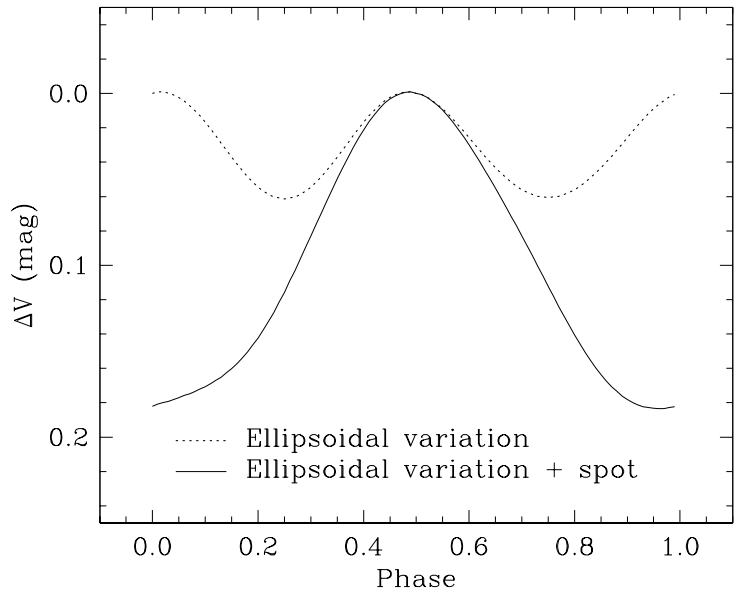}}
\caption{{\small Synthetic light curves for S\,1113, phased against
the binary orbit period. The upper curve shows expected light
variation due to the expected ellipsoidal shape of the primary, given
synchronous rotation. The lower curve shows the same light curve
combined with the photometric variation due to a spot at longitude
270$^{\circ}$. The resulting light curve is very similar to that seen
for S\,1113 (Fig.~\ref{s1063:1113phot}), and the ellipsoidal variations are no
longer evident.}} 
\label{s1063:ellips}
\end{figure}

The stars S\,1063 and S\,1113 in the field of the old open cluster M67 have attracted
attention for their location in the cluster color-magnitude diagram
roughly 1 magnitude below the cluster subgiant branch (Fig.~\ref{s1063:cmd}).
Comprehensive photometric
(optical and X-ray), spectroscopic, and astrometric studies have shown
them to be RS\,CVn-like systems with observed
three-dimensional motions that are the same as that of the cluster to high
precision.
Their location in the color-magnitude diagram has led us to describe these stars
as candidate ``sub-subgiant'' cluster members.

Specifically, S\,1063 is a single-lined spectroscopic binary with a
period of 18.4 days and an eccentric orbit ($e = 0.2$). The primary
effective temperature is 5000 K. Spectroscopic
surface gravity measurements indicate a subgiant 
with a surface gravity log g $\approx$ 3.5.
UBVIJHK colors indicate a reddening E(B-V) = 0.14 - 0.18, higher than
the cluster reddening of E(B-V)=0.032. As the entire Galactic reddening
in the direction of M67 is only 0.03 mag, the enhanced reddening of S1063
is likely not interstellar.
There is no evidence in the
near-infrared colors for warm circumstellar dust in this system.
The strong X-ray emission suggests rapid rotation of the primary,
although the measured $v\sin i$ is only 6 km s$^{-1}$.
Pseudosynchronous rotation at a period of 14.64 days is permitted by
existing observations, but
there is no evidence for photometric variability at this period.
Photometric variability is seen on longer timescales, with no evident
periodicity for periods less than 18 days.

S\,1113 is a double-lined spectroscopic binary with a period of 2.8
days, a circular orbit,
and a mass ratio of 0.7. Effective temperatures of 4800 K and 5500 K
are found for the primary and secondary stars, respectively.
The photometric properties of S1113 are well modeled by a binary with
a 0.9 $M_{\odot}$ main-sequence
secondary star and 1.3 $M_{\odot}$ a subgiant primary star (2.0 $R_{\odot}$) at a distance similar to M67.
However, this model does not easily explain the large projected rotation velocity (53 kms$^{-1}$) of the primary star.
Supersynchronous rotation of the primary
is required, in contrast to the short synchronization timescales, the circular orbit, and the observed variability
of the composite light with the orbital period.

Alternatively, mean densities for both stars are derived geometrically assuming a tidally
relaxed system. These densities indicate a 0.9 $M_{\odot}$ main-sequence secondary star and a 1.3 $M_{\odot}$, 
4.0 $R_{\odot}$ primary star that very nearly fills its Roche lobe.
In this model S1113 would be an RS CVn (possibly transferring mass) located behind
the cluster. However, the V flux ratio derived from this model 
differs by a factor 3 from the spectroscopically  measured flux ratio. In addition, the observed light curve shows
no evidence for asymmetry of the nearly-Roche-filling primary star. Thus
neither the photometric nor the geometric analyses are able to provide 
a fully consistent  model for the S1113 system.

The issue of cluster membership is pivotal for understanding both S1063
and S1113. Given their low apparent luminosity, the interpretation of
both binaries as background field RS CVn's is a natural solution. This interpretation
runs counter to the very close kinematic three-dimensional association of both binaries to the cluster.
Still, our statistical analyses indicate that the probability of one of the two
being a background binary is not entirely negligible, although both being background stars is
improbable. Equally important, we have not been
able to create model background binaries which reproduce all of the
observed properties of either S1063 or S1113.

Star clusters have long been recognized as dynamically active environments, particularly in the form of binary 
encounters with both single stars and other binaries. It would be no surprise if we were to find products of
such encounters that run counter to single-star evolutionary theory. Nor would it be a surprise if the products of such 
encounters were binary stars. S1063 and S1113 may be
recent products of such encounters, and as such may define 
alternative evolutionary tracks in the cluster environment. Regrettably we
have not yet been able to explain their low luminosities, although we conjecture
that their evolutionary histories include mass-transfer episodes, mergers, and/or 
dynamical stellar exchanges.

Albrow et al. (2001) have noted
stars in the globular cluster 47 Tuc which fall in a similar domain of the color-magnitude diagram as S1063 and 
S1113. Their finding of more examples of stars like S1063 and S1113 emphasizes the need for searches
for sub-subgiants in other clusters.

Finally, the primaries of both S1063 and S1113 are candidates
for future mass transfer. It is intriguing to consider these binaries
as progenitors of the blue stragglers, and particularly of
the short-period binary blue straggler F190 in M67.

\acknowledgments
We thank R. Davis, E. Horine, A. Milone
and J. Peters for acquiring many of the CfA spectra. We are grateful
to J. Orosz for assistance in computing the ellipsoidal variation
light curves. We thank Pierre Demarque and Sukyoung Yi for providing their
stellar
evolution models. We also appreciate a very helpful review by the referee
Roger Griffin. RDM and KS are supported by National Science Foundation
research grant AST 9731302. MvdB was supported by the Netherlands
Organization for Scientific Research.

\newpage
\begin{deluxetable}{rr}
\tablecolumns{2}
\tabletypesize{\footnotesize}
\tablewidth{0pt}
\tablecaption{Radial-velocity measurements for S\,1063 \label{s1063:rv1063t}}
\tablehead{ \colhead{HJD} & \colhead{$v_{\rm rad}$ (km s$^{-1}$)} }
\startdata
   2446899.6818 &     21.94 \\  
   7158.0132 	&     26.21 \\
   7198.7906 	&     44.67 \\
   7216.7438 	&     42.04 \\
   7226.7088 	&     17.03 \\
   7489.8904 	&     29.52 \\
   7493.8852 	&     46.04 \\
   7513.9289 	&     52.67 \\
   7515.8531 	&     53.39 \\
   7516.8802 	&     50.36 \\
   7519.8331 	&     21.04 \\
   7522.9810 	&     14.87 \\
   7524.0551 	&     17.48 \\
   7526.0601 	&     26.21 \\
   7543.8490 	&     23.10 \\
   7544.8949 	&     27.58 \\
   7545.8256 	&     33.04 \\
   7547.9579 	&     43.67 \\
   7549.9533 	&     52.55 \\
   7552.8487 	&     52.40 \\
   7555.8172 	&     31.66 \\
   7574.7462 	&     27.03 \\
   7579.7805 	&     19.95 \\
   7601.6697 	&     35.60 \\
   7608.6951 	&     52.30 \\
   7609.6639 	&     46.29 \\
   7610.6930 	&     36.26 \\
   7845.0409 	&     51.18 \\
\enddata
\end{deluxetable}

\begin{deluxetable}{lr}
\tablecolumns{2}
\tabletypesize{\footnotesize}
\tablewidth{0pt}
\tablecaption{Orbital elements for S\,1063 \label{s1063:rv1063el}}
\tablehead{ \colhead{} & \colhead{} }
\startdata
  $P$ (days)               & 18.396   $\pm$  0.005  \\
  $\gamma$ (km s$^{-1}$)   & 34.30  $\pm$  0.20     \\
  $K$ (km s$^{-1}$)        & 20.0  $\pm$  0.3       \\
  $e$                      & 0.206  $\pm$  0.014    \\
  $\omega$ (deg)           & 95  $\pm$  5           \\
  $T$ (2440000 +   days)       & 7482.19  $\pm$  0.22   \\
  $a_1 \sin i$  (AU)       & 0.0330  $\pm$  0.0006  \\
  $f(m)$  ($M_{\odot}$)    & 0.0143  $\pm$  0.0007  \\
                           &                        \\
  Number of observations   & 28                     \\
  $\sigma$ (km s$^{-1}$)   & 0.99                   \\ 
\enddata
\end{deluxetable}

\begin{deluxetable}{rrr}
\tablecolumns{3}
\tabletypesize{\footnotesize}
\tablewidth{0pt}
\tablecaption{Radial-velocity measurements for S\,1113 \label{s1063:rv1113t}}
\tablehead{ \colhead{HJD} & \colhead{$v_{\rm rad,p}$ (km s$^{-1}$)} & \colhead{$v_{\rm rad,s}$ (km s$^{-1}$)} 
}
\startdata
  2447579.7541 & $-$23.72 & 114.14   \\  
  47610.8114   & $-$24.26 & 117.76   \\
  47635.6961   & $-$4.62  & 87.84    \\
  47903.0130   & 92.82    & $-$48.87 \\
  48320.7433   & 93.49    & $-$48.93 \\
  48338.7448   & $-$18.47 & 103.52   \\
  48345.7097   & 74.53    & $-$22.31 \\
  48644.8675   & 64.94    & $-$5.16  \\
  49021.9789   & $-$11.59 & 105.53   \\
  49057.7946   & 83.51    & $-$38.66 \\
  49058.6665   & $-$9.15  & 103.29   \\
  49348.0117   & 83.93    & $-$36.01 \\
  49354.9585   & $-$5.61  & 86.68    \\
  49465.7519   & $-$16.13 & 103.53   \\
  49699.9173   & $-$26.20 & 112.09   \\
  50799.0251   & 73.97    & $-$28.94 \\
  50821.8463   & 92.94    & $-$53.73 \\
  50821.8681   & 94.91    & $-$53.55 \\
\enddata
\end{deluxetable}

\begin{deluxetable}{lrr}
\tablecolumns{3}
\tabletypesize{\footnotesize}
\tablewidth{0pt}
\tablecaption{Orbital elements for S\,1113. The primary is indicated
with '1', the secondary with '2'. \label{s1063:rv1113el}}
\tablehead{ \colhead{} & \colhead{} }
\startdata
   $P$ (days)             &  2.823094 $\pm$ 0.000014 \\
   $\gamma$ (km s$^{-1}$) &  33.4   $\pm$   0.4      \\
   $K_1$ (km s$^{-1}$)    &  60.6   $\pm$   0.9      \\
   $K_2$ (km s$^{-1}$)    &  86.2   $\pm$   0.6      \\
   $e$                    &  0                       \\
   $\omega$ (deg)          &  --                      \\
   $T_0$ (2440000 + days) & 8916.368 $\pm$ 0.004   \\
   $a_1 \sin i$  (AU)     &  0.0157   $\pm$   0.0003 \\
   $a_2 \sin i$  (AU)     &  0.0223   $\pm$   0.0003 \\
   $M_1 \sin ^3 i$ ($M_{\odot}$) & 0.544 $\pm$ 0.012 \\
   $M_2 \sin ^3 i$ ($M_{\odot}$) & 0.382 $\pm$ 0.012 \\
   $q$                    &  0.703   $\pm$   0.012   \\
                          &                          \\
   Number of observations &  18                      \\   
   $\sigma_1$  (km s$^{-1}$) &  3.15                 \\
   $\sigma_2$  (km s$^{-1}$) &  2.10                 \\
\enddata
\end{deluxetable}

\begin{deluxetable}{lrrrrrrrrr}
\tablecolumns{8}
\tabletypesize{\footnotesize}
\tablewidth{0pt}
\tablecaption{Photometric analyses of S\,1063. \label{s1063:phot}}
\tablehead{ \colhead{} & \colhead{$V$} & \colhead {$U-B$} &  \colhead{$B-V$} &
\colhead{$V-I$} &  \colhead{$J-H$} &  \colhead{$H-K$} &  \colhead{$L/L_{\odot}$}  & \colhead{$T_{\rm eff}$ 
(K)} &  \colhead{$R/R_{\odot}$} }
\startdata
   Observed  & 13.79    &   0.72 & 1.05 & 1.20 & 0.60 & 0.10 & &5000 & \\
   Intrinsic\tablenotemark{a} [E$(B-V) = 0.032$]   &        &   0.69 &  1.02 & 1.15 & & & &  & \\
   Model (1.37 M$_\odot$ 4 Gyr\tablenotemark{b})  &  &  0.84 &  1.02 & 1.01 & & &  & 4577 & \\
    &          &          &        & & & &\\
   Intrinsic\tablenotemark{a} [E$(B-V) = 0.18$]   &      &   0.55 &  0.87 & 0.90 & & &3.34 & 5000 & 2.4 \\
   Model (1.35 M$_\odot$ 4 Gyr\tablenotemark{b} )  &     &   0.56 & 0.87 & 0.90 & & &  &  4989 & \\
    &          &          &        & & & &\\
   Intrinsic\tablenotemark{a} [E$(B-V) = 0.14$]   &      &   0.60 &  0.91 & 0.97 & 0.55 & 0.07&3.28& 5000 & 2.4 
\\
   Model (1.35 M$_\odot$ 4 Gyr\tablenotemark{c} )  &     &   0.65 & 0.92 & 0.97 & 0.52 & 0.09 &  &  4921 & \\
    &          &          &        & & & &\\
   Model (2.3 M$_\odot$ 0.8 Gyr\tablenotemark{c} )  &     &   0.62 & 0.91 & 0.94 & 0.50 & 0.01 & 35.1  &  4990 
& 7.9 \\
\enddata
\tablenotetext{a}{Using extinction curve of Cardelli et al.\ (1989); luminosity and radius are derived assuming cluster 
membership.}
\tablenotetext{b}{Derived from Yale isochrones (solar metallicity, GDK transformation).}
\tablenotetext{c}{Derived from Yale isochrones (solar metallicity, LCD transformation).}
\end{deluxetable}
\begin{deluxetable}{lllllll}
\tablecolumns{7}
\tabletypesize{\footnotesize}
\tablewidth{0pt}
\tablecaption{Photometric analyses of S 1113. \label{s1063:1113decomp} }
\tablehead{ \colhead{} & \colhead{$V$} & \colhead{$U-B$} &  \colhead{$B-V$} &  \colhead{$L/L_{\odot}$}  & 
\colhead{$T_{\rm eff}$ (K)} &  \colhead{$R/R_{\odot}$} }
\startdata
   Observed Composite  & 13.77    & 0.52 &   1.01 & & & \\
   Primary    &   14.07  & 0.55\tablenotemark{a} &  1.07 \tablenotemark{a} & 2.0  & 4800  & 2.1  \\
   Secondary  &  15.30   & 0.45\tablenotemark{b} & 0.85 \tablenotemark{b} &   & 5500  &  \\
    &          &          &        & & & \\
   1.37 $M_{\odot}\tablenotemark{c}$ & & 0.88& 1.04    &      & 4534 &  \\
   0.90 $M_{\odot}\tablenotemark{c}$ & 5.673\tablenotemark{c} & 0.43&0.82     &     0.50  & 5311 &  0.83 \\   
\enddata
\tablenotetext{a}{Derived from observed composite colors and model secondary colors.}
\tablenotetext{b}{Derived from 0.90 M$_\odot$ model reddened by $E(B-V)=0.032$.}
\tablenotetext{c}{Intrinsic absolute magnitude M$_v$ and colors from Yale isochrone (solar metalicity, GDK 
transformation).}
\end{deluxetable}
\newpage
\end{document}